\documentclass[a4paper,12pt]{article}[]
\usepackage{epsf,float}

\newcommand{\for}{\qquad\mathrm{for}\qquad}

\newcommand{\Alpha}{\alpha\to -\infty}
\newcommand{\Nob}{``no boundary''}
\newcommand{\WDW}{Wheeler--DeWitt}
\newcommand{\KS}{Kantowski--Sachs}
\newcommand{\neun}{I$\!$X}
\newcommand{\reell}{\mbox{I$\!$R}}
\newcommand{\dis}{\displaystyle}

\newcommand{\ca}[1]{\ensuremath{\mathcal{#1}}}

\newlength{\abbildunterschriftlaenge}
\newlength{\abbildunterschrifteinrueck}
\begin{document}
\sloppy

\thispagestyle{empty}
\vspace*{-10mm}\hfill PITHA 94/64
\newline \vspace*{1mm}\hspace*{10cm}\hfill Juni 1995
\newline \vspace*{1mm}\hspace*{10cm}\hfill gr-qc/9412049

\vspace*{15mm}

\begin{center}
\Huge{\textbf{
Quantum Cosmology of \\
Kantowski--Sachs like Models
}}
\end{center}

\vspace*{5mm}

\begin{center}
\large{
Heinz--Dieter Conradi \\
Institute for Theoretical Physics E, RWTH Aachen \\
Sommerfeldstr. 26-28, 52056 Aachen, Germany \\
Email: conradi@physik.rwth-aachen.de \\
PACS--No.: 0460
}
\end{center}

\vspace*{20mm}
\begin{center}
\large{to appear in \\
\texttt{Classical and Quantum Gravity}}
\end{center}
\vspace*{15mm}

\begin{abstract}

  The \WDW\ equation for a class of \KS\ like models is
  completely solved. The generalized models include the \KS\
  model with cosmological constant and pressureless dust.
  Likewise contained is a joined model which consists of a \KS\
  cylinder inserted between two FRW half--spheres. The (second
  order) WKB approximation is exact for the wave functions of the
  complete set and this facilitates the product structure of the
  wave function for the joined model. In spite of the product
  structure the wave function can not be interpreted as admitting
  no correlations between the different regions.  This problem is
  due to the joining procedure and may therefore be present for
  all joined models.  Finally, the \underline symmetric
  \underline initial \underline condition (SIC) for the wave
  function is analyzed and compared with the \Nob\ condition. The
  consequences of the different boundary conditions for the arrow
  of time are briefly mentioned.

\end{abstract}

\newpage
\section{Introduction}
\typeout{Introduction}

Ever since the advent of canonical quantum gravity with DeWitt's
paper \cite{dewitt_67i}, minisuperspace models have played an
important part in the discussion. The popularity of this finite
dimensional models is based both on the otherwise tremendous
technical problems and the difficulty to find even classical
solutions for the general case. Although the relevance of
minisuperspace models is not clear, it is hoped that they shed
some light on the general theory and perhaps on some conceptual
issues, such as the problems of how to pose boundary conditions
and the arrow of time.

While far the most papers on minisuperspace quantization are
concerned with the isotropic FRW model, the \KS\ spacetime is the
most popular anisotropic model. The more realistic
Bianchi--\neun\ spacetime, unfortunately, has turned out to be
rather complicated; the classical model was a long time believed
to be chaotic and only a few exact solutions are known. For the
\KS\ model in comparison all solutions are known analytically
even if some particular types of matter are coupled to gravity.
It is
thus a good toy model for studying the influence of anisotropy in
quantum cosmology or more generally for models with additional
gravitational degrees of freedom to the one FRW degree. The \KS\
model has for example played a crucial role in Hawking's argument
why the arrow of time is asymmetric. In addition, since the \KS\
metric may describe the interior of black holes, it has been used
to discuss the entropy and the quantum states of black holes. The
last point has recently gained renewed interest in the
investigations of two--dimensional black holes following the
paper of Callan et.~al.  \cite{callan_etal_92}. This last point
mentions already one of the drawbacks of the quantization of the
\KS\ spacetime: namely that the \KS\ spacetime is not a spacetime
itself but only part of it. What does it mean to quantize only a
part of the system? This restriction holds for the vacuum model
but disappears if a perfect fluid is coupled to gravity.
Unfortunately, quantum cosmology with a perfect fluid is somewhat
artificial; however it is not sufficient to introduce a scalar
field in order to render the \KS\ metric complete.

In the next part of this paper some aspects of the classical \KS\
model are reviewed, thereby emphasizing certain exceptional
aspects. Moreover a joined model is introduced which is compact,
inhomogeneous and anisotropic. It consists of a \KS\ cylinder
which is inserted between two FRW half--spheres. In Sec. 3 a
whole class of \KS\ like models is quantized and a complete set
of exact solutions of the corresponding \WDW\ equation is
derived. In Sec. 4 a Hamiltonian formalism for the joined model
is derived which makes obvious that it can be quantized along the
lines of Sec. 3.  Special emphasize is given to the
interpretation of the resulting wave function. It turns out that
there is a hidden inconsistency in the quantization the
junction.

In order to render the \WDW\ equation physically
meaningful it is necessary to pose some sort of boundary
conditions at the wave functions.  In section 5 the \underline
symmetric \underline initial \underline condition (SIC) is
investigated for the \KS\ model. This initial condition is then
compared with the \Nob\ condition of Hartle and Hawking and the
consequences of the different boundary conditions on the arrow of
time are outlined.

\enlargethispage{\baselineskip}
Natural units $\hbar=c=G=1$ are used throughout.

\section{The classical \KS\ model}
\typeout{The classical Kantowski--Sachs model}

This section is intended to give a short review of the \KS\ model
by highlighting some of its features. The \KS\ spacetime which
was first investigated by Kompaneets and Chernov
\cite{kompaneets_chernov_64} and independently by Kantowski and
Sachs \cite{kantowski_sachs_66} combines spherical symmetry with
a translational symmetry in the ``radial'' direction. The
spacelike
hypersurfaces of constant times are therefore cylinders
\begin{equation}
  ds^2\ =\ z^2\,dr^2 + b^2\,d\Omega^2
\ .\label{metric}
\end{equation}
Here $b=b(t)$ is the surface measure of the two--spheres $S_2$
with metric $d\Omega^2$; $z=z(t)$ measures the spacelike distance
between the two--spheres and $r$ is the radial coordinate,
$r\in$\reell, or after compactification $r\in\ca{I}$, with
$\ca{I}$ an arbitrary interval of \reell. Neither $b$ nor $z$
depend on the angle variables of $S_2$ or on the radial
coordinate $r$.

It is important that the variables $b$ and $z$ are playing
different r\^oles since a cylinder is defined only by the
homogeneity of $b$ and would not be deformed by an inhomogeneous
$z$.  Actually an inhomogeneous $z=z(t,r)$ is even dynamically
consistent, e.~g. for matter in form of pressureless dust
\cite{ellis_67}, but this case is not considered in the remainder
of the paper.

There are two types of singularities in this geometry: cigarlike
singularities as $b\to0$ and disklike singularities as $z\to0$.
However, the curvature of the hypersurface $^3\ca{R}=2/b^2$ is
divergent for the cigarlike singularity only. An investigation of
the spacetime structure is thus unavoidable in order to analyze
the disklike singularity. It turns out that without matter it is
a coordinate singularity where the right--handed coordinate
system is changed into a left--handed one. Moreover it is an
indication for the imcompleteness of the vacuum \KS\ spacetime;
in the case of a vanishing cosmological constant it describes the
interior parts of the Kruskal spacetime\footnote{ Although a
  compactification of the radial axis does change the coordinate
  singularity into a topological singularity, it does not improve
  the incompleteness of the model.
}.
The model can be rendered
complete by adding matter e.~g. in form of pressureless dust,
since then the 4--curvature $^4\ca{R}$ diverges as the matter
density becomes infinite.

For this reason the case of pressureless dust is considered here,
although the quantization of this model is somewhat artificial. A
scalar field as a more realistic matter field is not considered.
In the first place because the differential equations with a
massive scalar field are much more difficult and neither the
classical nor the quantized ones have yet been analytically
solved. Second, the model *seems to remain* incomplete because as $z\to0$
the scalar field behaves like an effective cosmological constant;
that means, for $z\to0$ the model with scalar field *would be* as
incomplete as the vacuum one. Furthermore the construction of
the joined model at the end of this section is impossible with a
scalar field.  However, the scalar field can, at least for
minisuperspace models, be conveniently used in order to describe
a dust field by considering the limiting case when
$8\pi\rho=\dot\phi^2/N^2 + m^2\phi^2\to const$ and simultaneously
$p=\dot\phi^2/N^2 - m^2\phi^2\to0$. The equation of motion for
the scalar
field
\begin{equation}
  \frac 1N\left(\frac{\dot{\phi}}{N}\right)^. +
  \frac{1}{N^2}\frac{\dot{v}}{v} + m^2\phi^2 = 0\ ,
\end{equation}
where $v=zb^2$ is the volume measure of the hypersurface, proves
this possibility if the second term can be neglected (which is
fulfilled for great volumes) for $\phi=\sqrt{8\pi\rho}/m \sin
m\tau$, with $d\tau=Ndt$.  The dust is then described by a
parameter instead of an variable in the Lagrangian, c.~f.  Eq.
(\ref{kslagrange}). This approximation breaks down, of course,
for vanishing volumes. However, for small volumes the scalar
field behaves as an effective cosmological constant. It may
therefore be more realistically to consider the regions for small
and for great volumes separately. It should be noted that this
approach which treats dust as a mere parameter is essentially
equivalent to a more sophisticated approach starting point of
which is a Lagrangian for the dust degrees of freedom, see e.~g.
\cite{brown_kuchar_94a} and the literature cited therein. This
is, because the latter approach leads in the homogeneous models
to a Lagrangian (Hamiltonian) which consists of only one term
containing a cyclic momentum as the only remaining variable.

The Lagrangian of the \KS\ model is
\begin{equation}
   L(z,b)\ =\
  \frac{\ca{I}}{2}\left( -\frac{2b\dot{b}\dot{z}}{N} -
  \frac{z\dot{b}^2}{N} + N\left( z - z_{\mathrm{m}} -
  \Lambda zb^2 \right)\right)
\label{kslagrange}
\end{equation}
where $\ca{I}$ is the compactification interval, $\Lambda$ is a
cosmological constant and $z_{\mathrm{m}}=\rho zb^2=const$ is the
dust potential (analogous to $a_{\mathrm{m}}=\rho a^3$ in the FRW
model). Considering the above comment on dust in quantum
cosmology one may thus approximate the scalar field for small
volumes by $\Lambda\not=0$ and $z_{\mathrm{m}}=0$ and for large
volumes by $\Lambda=0$ and $z_{\mathrm{m}}\not=0$.  If a two
dimensional pseudosphere or plane is considered instead of the
two--sphere in (\ref{metric}) a different sign of the first term
of the potential will result: $-z$ respectively $0$ instead of
$+z$.  These models are special cases of the generalized \KS\
model which is introduced and quantized in Sec.~3.

The Legendre transformation with the canonical momenta
\begin{equation}
  \pi_b = -\frac{\ca{I}(zb)\dot{}}{N} {\rm \qquad and\qquad}
  \pi_z = -\frac{\ca{I}b\dot{b}}{N}
\end{equation}
leads to the Hamiltonian constraint
\begin{equation}
  z\dot{b}^2\ +\ 2\dot{z}b\dot{b}\ +\ N^2 \left(
  z\ -\ z_{\mathrm{m}}  - \Lambda zb^2 \right)\ =\ 0
\ ,\label{hamilton}
\end{equation}
written in the configuration
variables and their velocities.
The equations of motion are obtained by taking the Poisson
brackets with the Hamiltonian,
\begin{eqnarray}
 && \frac{\dot{b}^2}{N^2} + 1 - \frac{b_{\mathrm{m}}}{b} -
  \frac 13 \Lambda b^2 \ =\ 0
\ ,\label{eqb}
\\
 && b\ddot{z} + \dot{b}\dot{z} + \ddot{b}z - (zb)\dot{}\
  \frac{\dot{N}}{N} - N^2zb\Lambda
  \ =\ 0 \ ,
\label{eqz}
\end{eqnarray}
where the equation of motion for $b(t)$ has already been
integrated. Eq.  (\ref{eqb}) therefore contains $b_{\mathrm{m}}$
as an arbitrary constant of motion; for a vanishing cosmological
constant $b_{\mathrm{m}}$ is the maximum of $b(t)$.  Although
this equation is identical to the equation of motion for the
scale factor $a(t)$ of the closed FRW model, in contrast to that
model the matter content of the \KS\ model does {\it not} fix the
constant of integration $b_{\mathrm{m}}$. Instead, the matter
content fixes the constant of integration for the $z$ dynamics as
already indicated by the notation: $\rho
zb^2:=z_{\mathrm{m}}=z(t_{\mathrm{m}})$ where $t_{\mathrm{m}}$ is
defined by $b(t_{\mathrm{m}})=b_{\mathrm{m}}$. It is worth noting
that Eq.~(\ref{eqb}) provides one with an expression for the
classical range of the variable $b$: $b_{\mathrm{m}}>b(1-\frac
13\Lambda b^2)$.  Since the equation of motion for $b(t)$ is
independent of $z(t)$, these equations confirm that an
inhomogeneous $z(t,r)$ might be dynamically consistent. A
rigorous proof must, however, use the equations of motion for the
general spherically symmetric model. There it follows directly
since, in Kucha\v{r}'s notation \cite{kuchar_94a} (there are two
fields $\Lambda(t,r)$ and $R(t,r)$ with $\Lambda(t,r)\to z(t,r)$
and $R(t,r)\to b(t)$ for the \KS\ model), terms which contain
$\Lambda'$ do always contain $R'$ too and do thus not appear in
the \KS\ model.

In order to solve the equations of motion the lapse function has
to be fixed. The gauge $N=b$ has the advantage of covering the
whole manifold while, for example, the gauge $b=t$ covers only
half of it. For a vanishing cosmological constant, the explicit
solutions for the gauge $N(t)=b(t)$
are
\begin{eqnarray}
  b(t)\ &=&\ b_{\mathrm{m}} \sin^2\left(\frac{t}{2}\right)
\ ,\label{classical} \\
  z(t)\ &=&\ K\cot\left(\frac{t}{2}\right) +
  z_{\mathrm{m}}\left( 1 -
  \frac{t}{2} \cot\left(\frac{t}{2}\right) \right)
\ ,\nonumber
\end{eqnarray}
where the initial time has been set to zero and where $K$ is an
arbitrary constant of integration. For an illustration see the
left diagram in Fig.~1 where the model is depicted in
Kruskal--type coordinates.  It is worthwhile to note that in the
vacuum model the constant of integration $K$ is meaningless due
to a possible redefinition of the radial coordinate $r\to Kr$.
Solutions with different values of $K$ are thus identical. The
inhomogeneity of $z(r,t)$ is due to either inhomogeneous dust
$z_{\mathrm{m}}(r)$ or to an inhomogeneous constant of
integration $K(r)$. Unlike the FRW model, the \KS\ model is not
time--symmetric with respect to $b_{\mathrm{m}}$ since $z(t)$ is
not.

For the case of $z_{\mathrm{m}}=0$ but $\Lambda\not=0$ one
obtains a similar set of solutions which admits the
exceptional solution of a constant $b(t)$:
$b=\sqrt{1/\Lambda}$ and $z=e^t$ (again in the gauge $N=b$).

Since both the \KS\ and the FRW hypersurface are spherically
symmetric, it is possible to construct a new spherically
symmetric 3--geometry by \textit{joining} the respective
hypersurfaces. In Fig.~2 such a hypersurface is visualized.  In
order to be dynamically consistent, however, the dynamics of the
surface measure of the two--sphere, the metric at fixed $r$, has
to be identical on both sides of the junction. This is trivially
fulfilled in this case, provided the constants of integration are
equal: $b_{\mathrm{m}}=a_{\mathrm{m}}$. That means, the matter
content of the FRW part fixes the constant of integration of the
\KS\ part of the spacetime. In addition the metric of the
hypersurface has to be smooth (more precisely: in the notation by
Kucha\v{r} one has to require $R$ and $R'/\Lambda$ to be smooth
in order to obtain a well defined Hamiltonian).  This requires
the junction to take place at the equator of the FRW 3--sphere
which is thus cut into halves. It is therefore even possible to
split the FRW 3--sphere by inserting a \KS\ cylinder of arbitrary
length; this has no influence on the FRW dynamics. For an
illustration see the right diagram in Fig.~1 where the model is
represented in Kruskal--type coordinates. A Hamiltonian
formulation and quantization of this compact, anisotropic and
inhomogeneous model is given in Sec.~4. The joining procedure is
impossible with a scalar field as matter source. Techically, this
is because then the $b$--dynamics is not decoupled from the
$z$--dynamics. This means physically that the scalar field can
not be contained in one region: it leaks into the other regions,
thereby displaying the inhomogeneity of the model.

\begin{figure}[H]
\parbox{75mm}{\epsfxsize=75mm
  \epsffile{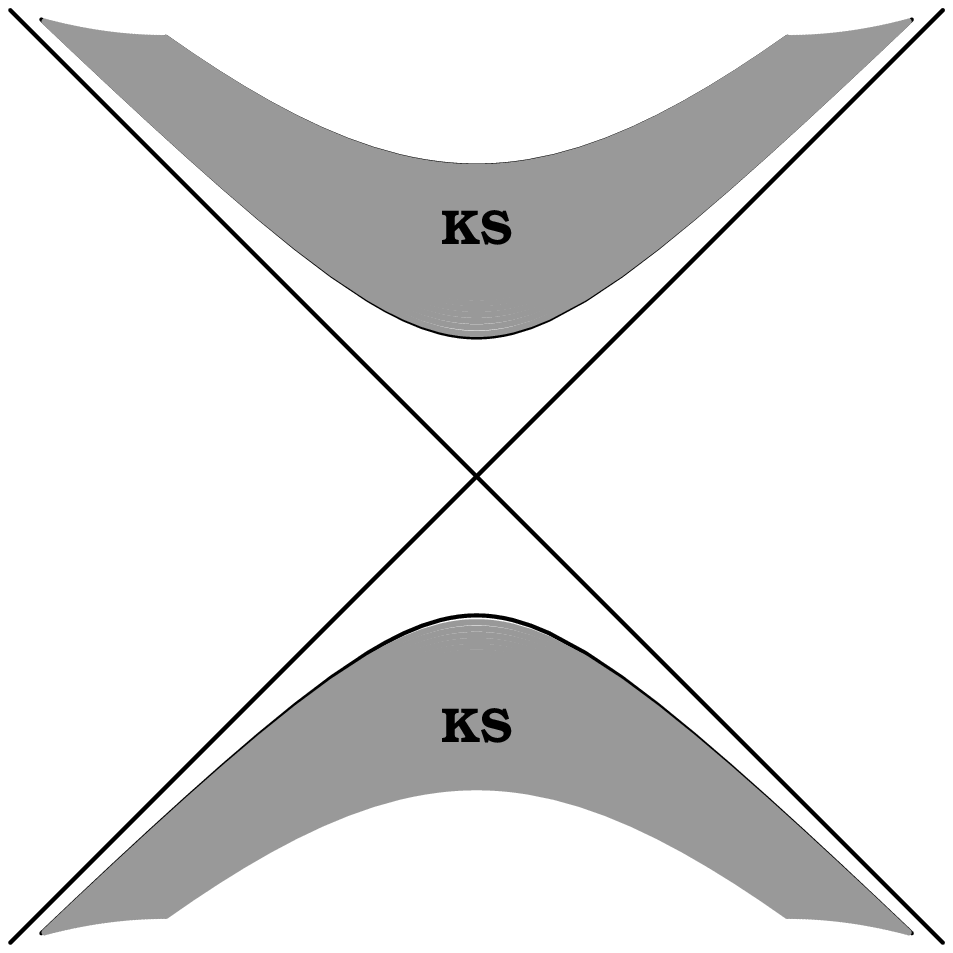}
}\hfill
\parbox{75mm}{\epsfxsize=75mm
  \epsffile{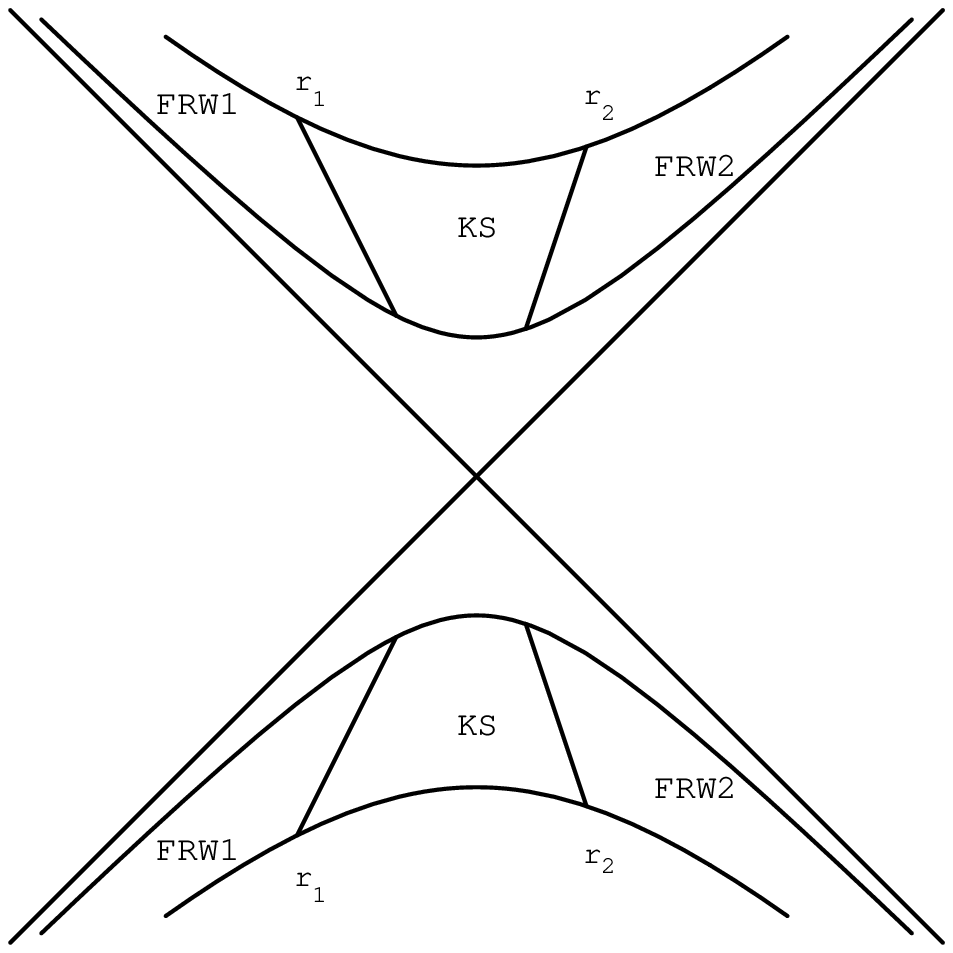}
}
\hspace*{\abbildunterschrifteinrueck}
\parbox{\abbildunterschriftlaenge}{
  {\label{kruskaldia} \footnotesize
  Fig.~1:
    The left diagram shows the dust filled \KS\ model in Kruskal--like
    coordinates.
    In the limit of vanishing dust the limiting
    hyperbolas degenerate into the usual lightlike event horizon of the
    Schwarzschild black hole. In that case it is possible to continue the
    \KS\ model into the outer regions of the Kruskal diagram. This is
    impossible in the dust case since there the hyperbolas build a spacelike
    borderline; the complete spacetime is represented
    by the shaded regions only. It is
    possible to continue paths (as e.~g. the $r=const$
    geodesics which are represented in this diagram by straight lines
    through the origin) from the lower into the upper \KS\ region. The
    right diagram shows as an example those geodesics which constitute the
    junction to the FRW regions of the joined model. In order to make
    graphical explicit that there is no crossing of the geodesics, right and
    left have been interchanged in the upper (lower) region.
}}
\end{figure}

\begin{figure}[H]
\center{\parbox{110mm}{\epsfxsize=110mm
  \epsffile{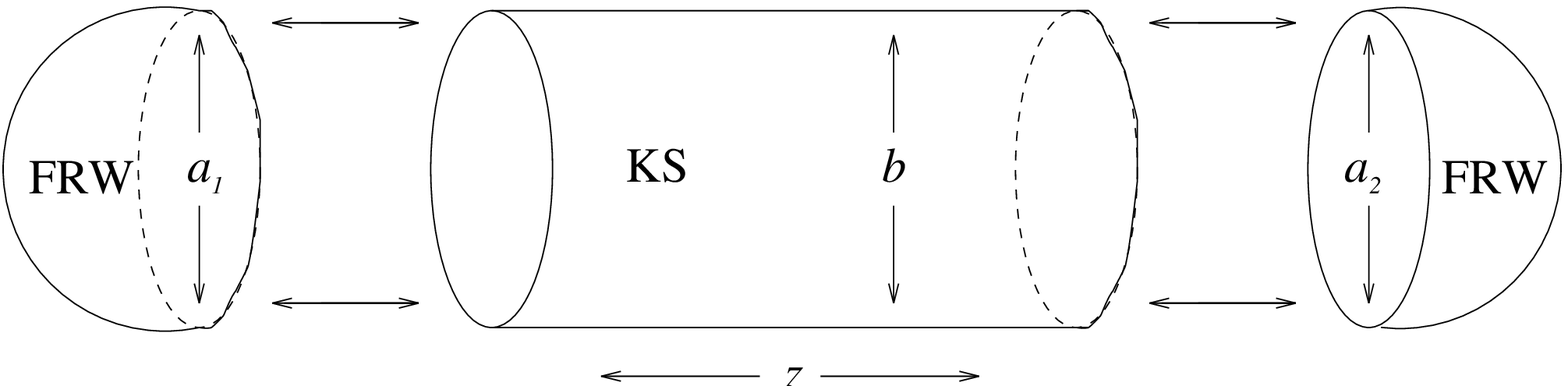}
}}
\hspace*{\abbildunterschrifteinrueck}
\parbox{\abbildunterschriftlaenge}{
  {\label{bild} \footnotesize
  \ \newline
  Fig.~2:
  The diagram shows a hypersurface of constant time for the joined model.
  Every $S_1$--circle represents a $S_2$--sphere.
  The arrows indicate the joining procedure which enforces $a_1=b=a_2$.
}}
\end{figure}

\section{Solving the \WDW\ equation}
\typeout{Solving the Wheeler--DeWitt equation}

As usual, the \KS\ model is quantized following Dirac's
quantization scheme by turning the variables $b$, $z$, $\pi_b$
and $\pi_z$ into operators which fulfill the standard commutation
relations and the Hamiltonian constraint (\ref{hamilton}) becomes
an operator which annihilates the physical states: $H=0\ \to\
\hat H\Psi=0$. In the field representation this equation is
called the \WDW\ equation. For the \KS\ model it takes the form
%
\begin{eqnarray}
&&
\hspace*{-9mm}
  -2\ca{I}zb^2\hat{H}\Psi=
  \Bigg[ z^2\partial^2_z + k_1z\partial_z -
  2zb\partial_b\partial_z
  + \ca{I}^2zb^2 \left( zf_1(b) + f_2(b)  \right)\Bigg]
  \Psi(b,z) = 0
\ ,\label{wdw} \\
  & & \mathrm{\hspace*{20mm} with\qquad}
  f_1=1-\Lambda b^2 \mathrm{\quad and\quad}
  f_2=-z_{\mathrm{m}}
\ ,\label{special}
\end{eqnarray}
where $\partial_z$ and $\partial_b$ are the partial derivatives
with respect to $z$ and $b$, respectively. Factor ordering is
partially left open as indicated by the arbitrary parameter
$k_1$. This one--parameter family of orderings includes the
important Laplace--Beltrami ordering with $k_1=1/2$.

The particular form (\ref{wdw}) of the \WDW\ operator with the
two functions $f_i(b)$ has been chosen because it is possible to
solve the equation exactly for arbitrary functions $f_i(b)$. This
general form of the Hamiltonian does not only include the
ordinary \KS\ model with dust and cosmological constant but also
the joined model (see Sec. 4) and the rotational symmetric
Bianchi--I and Bianchi--I$\!$I$\!$I models which result when the
two--spheres in the metric~(\ref{metric}) are replaced by
two--dimensional planes or pseudospheres, respectively. In the
literature one finds several sets of exact solutions for the
vacuum model, see e.~g.
\cite{louko_vachaspati_89,uglum_92,ryan_72}.  One of these is the
general solution (\ref{psi}) specialized for the vacuum case
\cite{louko_vachaspati_89}. The other two sets of solutions
utilize transformations which seperate the kinetic part of the
\WDW\ equation and at the same time keep the potential part
simple.

In order to solve the \WDW\ (\ref{wdw}) equation it is convenient
to introduce the operator
\begin{equation}
  2\ca{I}^2\hat{M}\ =\ -\frac 1b\partial_z{}^2 + \ca{I}^2\int^bf_1(b')db'
\label{M}
\end{equation}
(for the special case (\ref{special}) this simplifies to
$2\ca{I}^2\hat{M} = - (1/b)\partial_z{}^2 + \ca{I}^2(b-1/3
\Lambda b^3)$).  This operator is known from the general
spherical symmetric model where $\hat M(r)$ is interpreted as the
gravitating mass inside the radius $r$. (For an explicit account
of the general spherical model see
\cite{thiemann_kastrup_93a,thiemann_kastrup_94b,kuchar_94a}). One
can use this operator in the \KS\ model since $\hat M$ is the
first time--integral of the equation of motion for $b(t)$ (that
means, Eq.~(\ref{M}) is identical to the operator version of
Eq.~(\ref{eqb}) with $\hat M=2\hat b_{\mathrm{m}}$), although
neither the interpretation as gravitational mass nor its usual
derivation (which makes use of the supermomentum density) is
valid. This is true also for the general spherically symmetric
models (assuming vanishing shift--function). It is essential that
the new operator $\hat M$ has a vanishing commutator with the \WDW\ operator
\begin{equation}
  \left[ \hat M,\ \hat H\right]\ =\ 0
\ .
\end{equation}
One can thus find simultaneous eigenfunctions for both operators
and $\hat M$ can be interpreted as formal observable. The
eigenvalue equation $\hat M\psi_M=M\psi_M$ can, in
contrast to the \WDW\ equation (\ref{wdw}), easily be solved
\begin{eqnarray}
  &&\psi_{M}(z,b)\ =\
  h(b)\exp\left\{\pm i\ca{I} zW_M(b)\right\}
\ ,\label{psim} \\
  &&\hspace*{8mm}\mathrm{with}\qquad
  W_M^2(b) = 2Mb - b\int^bf_1(b')db'
\ .\nonumber
\end{eqnarray}
The arbitrary function $h(b)$ is then fixed by requiring $\psi_M$
to solve the \WDW\
equation. The resulting differential equation for $h(b)$ is again simple,
\begin{equation}
  \frac{h'(b)}{h(b)}\ =\ -\frac{\partial_bW_M(b)}{W_M(b)} +
  \frac{k_1}{2b} - \frac{i\ca{I}}{2G}\frac{bf_2(b)}{W_M(b)} \ .
\end{equation}
This gives (up to an arbitrary constant)
\begin{equation}
  \Psi_{_M}(z,b)\ =\ \frac{\sqrt{b^{k_1}}}{W_M(b)} \ \exp
  \left\{ \pm i\ca{I}
  \left( zW_M(b) - \frac 12 \int^b \frac{b'f_2(b')}{W_M(b')}db'
  \right) \right\}
\label{psi}
\end{equation}
as a one--parameter set of exact solutions for the \WDW\
equation.  In the particular model with $f_1(b)=1-\Lambda b^2$
and $f_2(b)=-z_{\mathrm{m}}$, the
wave function is thus
%
\begin{eqnarray}
  \hspace*{-5mm}
  \Psi_{_M}(z,b) &=& \sqrt{\frac{b^{k_1-2}}{
  \Lambda b^2/3 + 2M/b - 1 }}
  \exp\left\{\pm i\ca{I}\left( zb\sqrt{
  \frac{\Lambda b^2}{3} + \frac{2M}{b} - 1} + g(b) \right)\right\}
\label{psiks} \\
& & \mathrm{with} \qquad
  g(b)\ =\ \frac{z_{\mathrm{m}}}{2}\int^b
  \sqrt{\displaystyle\frac{b'}{\Lambda b'^3/3 + 2M - b'}}\ db'
\ .
\end{eqnarray}
The integral $g(b)$ can be solved analytically in special
cases. For $M = 0$, that is for large values of $b$, one gets
\begin{equation}
  g_{_{\Lambda}}(b)\ =\
  \frac{z_{\mathrm{m}}}{2}\sqrt{\frac{3}{\Lambda}}\ln
  \left( b + \sqrt{b^2 - \frac{3}{\Lambda}}\right)
\ ,\label{glambda}
\end{equation}
while for vanishing $\Lambda$, that is for small values of $b$,
one gets
\begin{equation}
  g_{_M}(b)\ =\ -\frac{z_{\mathrm{m}}}{2}\left(
  b\sqrt{\frac{2M}{b} - 1} + M\arcsin\left(\frac{M - b}{M}\right)
\right) \ .
\end{equation}

The solution for the model which approximates the scalar field at
vanishing $z_{\mathrm{m}}$ for small volumes and at vanishing
$\Lambda$ for great volumes is therefore given by Eq.
(\ref{psi}) with vanishing $g(b)$ and with $g(b)$ given by Eq.
(\ref{glambda}), respectively. These solutions have, of course,
to be matched at that volume where the character of the equation
changes.

These exact solutions (\ref{psi}) of the \WDW\ equation have the
peculiar structure that
\begin{equation}
  \Psi_{_M}=De^{\pm iS_{_M}}\ ,
\end{equation}
where $S_{_M}$ is a solution of the Hamilton--Jacobi equation and
$D$ is the (first order) WKB prefactor.  One may interprete these
solutions as a complete, one parameter family of exact solutions;
complete in the sense that $S_{_M}$ is a complete Integral of the
Hamilton--Jacobi equation. The prefactor $D$ does not depend on
the $f_2$--part of the potential; that means it does not depend
on the dust content. The considered factor ordering ambiguity has
completely been taken care of by the single factor
$b^{\frac{k_1}{2}}$.  This specific structure of the solutions
has important consequences in the next sections.  The solutions
are exact in spite of the divergency at the borderline of the
classically allowed region. It is possible to get rid of this
apparent divergency. Consider first the second term in the
exponend.  This is an integral with a lower limit of integration
which has not yet been fixed. The integral can therefore be
forced to vanish at the classical borderline by requiring the
integration to start just there (provided there is only one).
Since thus only the first part of the Hamilton--Jacobi function
is concerned one can simply consider the particular superposition
of the solution which is proportional to the sine.  At the
borderline the wave function behave then as $(1/x)\sin x$ and is
regular.

An exact WKB solution has recently been found for the much more
complicated Bianchi--\neun\ model
\cite{moncrief_ryan_91,ryan_93}.  That solution even has a
constant prefactor but it requires a rather artificial factor
ordering and it is a solitary solution instead of a complete set.
It can be obtained by transforming the trivial solution of the
\WDW\ equation in Ashtekars variables into geometrodynamical
variables. In this sense it is a ``ground state'' solution.

\section{The joined model}
\typeout{The joined model}

In this section a Hamiltonian formulation for the joined model
which consists of two FRW half--spheres and a connecting \KS\
cylinder is derived and then the model is quantized along the
usual lines. Without a Hamiltonian formalism for the
\textit{whole} model it is only possible to join the WKB
solutions of different sections as, e.~g., in
\cite{nambu_sasaki_88}, but not to go beyond this approximation.
However, it becomes apparent at the end that it is not possible
to really circumvent the difficulties of the junction procedure
with this approach.

The Lagrangian for the whole model is
\begin{equation}
  S(a,z,b)\ =\
  \int\int_0^{\frac{\pi}{2}}\ca{L}_{_{FRW_1}} d\chi dt +
  \int\int_{r_1}^{r_2}\ca{L}_{_{KS}} dr dt +
  \int\int_{\frac{\pi}{2}}^{\pi}\ca{L}_{_{FRW_2}} d\chi dt
\ ,\label{prelagrange}
\end{equation}
where $\chi$ is the third angle variable of the FRW three--sphere
and it has already been implied that the \KS\ model must join the
FRW model at the equator of the FRW spheres. The length of the
\KS\ cylinder is not restricted; that is, $r_1$ and $r_2$ are
arbitrary. All together, the following junction conditions have
to be fulfilled
for the joined model
\begin{eqnarray}
  &&b\ =\ a_i \sin\chi_i \ ,
\nonumber\\
  &&\sin\chi_i\ =\ 1 \ ,   \hspace*{35mm} i = 1,\ 2
\label{conditions} \\
  && b_{\mathrm{m}}\ =\ a_{m_i} = a^3\rho_{FRW_i}
\ , \nonumber
\end{eqnarray}
in order to built a dynamically consistent smooth hypersurface.
The $a_{m_i}$ are fixed by the dust contents of the respective
FRW regions.  The last condition thus means in particular that
the dust density of both FRW regions has to be identical; but
there is no requirement concerning the dust content of the \KS\
region. Inserting these conditions into the Lagrangian results
in a Lagrangian which
describes the whole model
\begin{eqnarray}
  L_{_{Ges}}(z,b)\ &=&\
  \frac{3\ca{I}_1}{2}\left( -\frac{b\dot{b}^2}{N} + N
  \left( b - a_{\mathrm{m}} -
  \frac 13 \Lambda b^3 \right)\right)\
\nonumber\\
  & &+\ \frac{\ca{I}_2}{2}\left( -\frac{2b\dot{b}\dot{z}}{N} -
  \frac{z\dot{b}^2}{N} + N\left( z - z_{\mathrm{m}} -
   \Lambda zb^2 \right)\right)
\ .\label{lagrange}
\end{eqnarray}
Here $b$ is both a \KS\ variable and (two times) a FRW variable
while $z$ is a genuine \KS\ variable.
$\ca{I}_1=\int_{_0}^{^{\pi}}sin^2\chi\,d\chi$ and
$\ca{I}_2=\int_{_{r_1}}^{^{r_2}}\,dr$ are the volume elements for
the FRW model and for the \KS\ model, respectively. This
Lagrangian can now be used to construct the Hamiltonian for the
joined model. It is important to notice that it is not possible
to incorporate the condition $a_{\mathrm{m}}=b_{\mathrm{m}}$ into
the Lagrangian since $b_{\mathrm{m}}$ does not occur explicitly.
Therefore still one junction condition has to be imposed.

The Legendre transformation with the canonical momenta
\begin{equation}
  \pi_z\ =\ -\frac{\ca{I}_2}{N}b\dot b \qquad\qquad
  \pi_b\ =\ -\frac{3\ca{I}_1}{N}b\dot b -\frac{\ca{I}_2}{N}(zb)\dot{}
\label{pifrwks}
\end{equation}
then leads to the Hamiltonian
\begin{eqnarray}
  \hspace*{-25mm}
  H\ &=&\
  \frac{1}{2\ca{I}_2 zb^2}\Bigg\{  z^2\pi_z^2 - 2zb\pi_z\pi_b -
  \ca{I}_2{}^2zb^2\left( z - z_{\mathrm{m}} -
  \Lambda zb^2 \right) +
\nonumber\\
  &&\hspace*{30mm} +\
  3\mbox{$\frac{\ca{I}_1}{\ca{I}_2}$}zb\left( \pi_z^2 -
  \ca{I}_2{}^2 \left(
  b^2 - ba_{\mathrm{m}} - \frac 13\Lambda b^4\right)\right)
  \Bigg\}\ =\ 0
\label{joinhamilton}
\end{eqnarray}
which is constrained to vanish. However, due to the neglected
junction condition this Hamiltonian alone does not reproduce the
correct equations of motion. It has to be accompanied by another
constraint which enforces $a_{\mathrm{m}}=b_{\mathrm{m}}$. It is
here chosen to be
\begin{equation}
  \ca{I}_2{}^2\left( M - \frac{a_{\mathrm{m}}}{2}\right)\ :=
  \frac{1}{2b}\pi_z{}^2 + \frac{\ca{I}_2{}^2}{2}b \left( 1 -
  \frac{\Lambda}{3}b^2 - \frac{a_{\mathrm{m}}}{b} \right) \ =\ 0
\ .\label{mjoin}
\end{equation}
The field $M$ which is defined as in the last section is fixed by
this constraint to the value $a_{\mathrm{m}}/2$.  The full
Hamiltonian $\ca{H}_{_{full}}$ leading to the correct equations
of motion is therefore given by
\begin{equation}
  \ca{H}_{_{full}}\ :=\ N_{_{H}}H + N_{_{M}}M
\label{fullham}
\end{equation}
with the two Lagrange multiplicators $N_{_{H}}$ and $N_{_{M}}$.
This procedure is consistent due to the vanishing of the Poisson
bracket of the two constraints
\begin{equation}
  \left\{ H,\ M\right\}\ =\ 0
\ ,
\end{equation}
that is $H=0$ and $M=0$ are first class constraints.  This
relation holds for the corresponding operators too, if for
example the simplest ordering for the operator $\hat{M}$ is
chosen and the Laplace--Beltrami factor ordering is chosen for
the \WDW\ equation which then reads
\begin{eqnarray}
\hspace*{-20mm}
  \hat{H}\Psi\ &=&\ -\frac{1}{2\ca{I}_2zb^2} \Bigg[ \left(z^2 +
  3\mbox{$\frac{\ca{I}_1}{\ca{I}_2}$}\right)
  \partial_z^2 + z\partial_z - 2zb\partial_z\partial_b
\nonumber\\
  &&\qquad +\
  \ca{I}_2{}^2 zb^2\left( z - z_{\mathrm{m}} - \Lambda zb^2 +
  3\mbox{$\frac{\ca{I}_1}{\ca{I}_2}$}\left( b - a_{\mathrm{m}} -
  \frac 13\Lambda b^3 \right)\right)
  \Bigg] \Psi = 0
\ . \label{joinwdw}
\end{eqnarray}
It is not surprising that these two operators commute because
$\hat{H}$ can be transformed into the generalized \KS\ model
(\ref{wdw}) by
\begin{equation}
  \bar{b}:=b \qquad\mathrm{and}\qquad
  \bar{z}:=z+\mbox{$\frac{\ca{I}_1}{\ca{I}_2}$}b \ .\label{trafo}
\end{equation}
In the notation of the last section $k_1=1/2$ and
\begin{equation}
  f_1(b)\ =\ 1 - \Lambda b^2
  \qquad\mathrm{and}\qquad
  f_2(b)\ =\ - \left(z_{\mathrm{m}} +
  \mbox{$\frac{\ca{I}_1}{\ca{I}_2}$}
  3b_{\mathrm{m}}\right) +
  2\mbox{$\frac{\ca{I}_1}{\ca{I}_2}$} b
\ .
\end{equation}
Only $f_2$ thus differs from the \WDW\ equation of the pure \KS\
model. $\hat{M}$ is not altered by the transformation.

Instead of one equation as in Sec. 3, two
differential equations have now to be satisfied
\begin{equation}
  \hat{H}\Psi\ =\ 0      \qquad\mathrm{and}\qquad
  \left( 2\hat{M} - a_{\mathrm{m}} \right)\Psi\ =\ 0
\label{twoeq}
\end{equation}
which can be viewed as a consequence of the inhomogeneity of the
model.  This does not pose any problems since in Sec. 3 the
solutions of the \WDW\ equation $\hat{H}\Psi=0$ were actually
found by using $\hat{M}\Psi=M\Psi$.  The sole consequence of the
second equation is the uniqueness of the
solution (up to the sign in the exponent)
\begin{eqnarray}
\hspace*{-20mm}
  \Psi_{_{\mathrm{join}}}(\bar{b},\bar{z}) &=& \frac{1}{
  \sqrt{\frac{\Lambda \bar{b}^3}{3} + a_{\mathrm{m}} - \bar{b}} }
  \exp\left\{\pm i\ca{I}_2\left( \bar{z}\bar{b}\sqrt{
  \frac{\Lambda \bar{b}^2}{3} + \frac{a_{\mathrm{m}}}{\bar{b}} - 1} +
  g(\bar{b})
  \right)\right\}
\label{psijoin} \\
& & \hspace*{-10mm}\mathrm{with} \qquad\quad
  g(\bar{b})\ =\ \frac 12{\int}^{\bar{b}} \ \dis
  \frac{z_{\mathrm{m}} + 3\frac{\ca{I}_1}{\ca{I}_2}a_{\mathrm{m}} -
  2\frac{\ca{I}_1}{\ca{I}_2}\bar{b}'}{
  \sqrt{\Lambda \bar{b}'^2/3 + a_{\mathrm{m}}/\bar{b}' - 1}}
  \ d\bar{b}'
\end{eqnarray}
since the eigenvalue of $\hat{M}$ is now fixed by the second
constraint.  It is of course possible to make the same
distinction between small volumes with vanishing dust and large
volumes with vanishing cosmological constant as in the last
section.

The WKB structure ($\Psi=De^{\pm iS}$) of this exact solution
allows one to write the wave function in another form. This is
most directly seen in the case of a vanishing cosmological
constant, but holds in the general case, too. The
Hamilton--Jacobi function $S$ is the sum of the Hamilton--Jacobi
functions of the separate regions $S=S_{_{KS}}+S_{_{FRW}}$
written in the original variables $b$ and~$z$
\begin{eqnarray}
  S &=& \ca{I}_2\left(
  b\left(z-\frac{z_{\mathrm{m}}}{2}\right)
  \sqrt{\frac{a_{\mathrm{m}}}{b}-1} -
  \frac{z_{\mathrm{m}}a_{\mathrm{m}}}{4}
  \arcsin\left(\frac{a_{\mathrm{m}}-2b}{a_{\mathrm{m}}}\right)
  \right)
\nonumber \\
& &
  +3\ca{I}_1\left(
  \frac{b^2}{4}\left(2-\frac{a_{\mathrm{m}}}{b}\right)
  \sqrt{\frac{a_{\mathrm{m}}}{b}-1} -
  \frac{a_{\mathrm{m}}^2}{8}\arcsin\left(
  \frac{a_{\mathrm{m}}-2b}{a_{\mathrm{m}}}\right) \right)
\ .\label{hjjoin}
\end{eqnarray}
Furthermore, the WKB prefactor (for the whole model) is here
simply the prefactor of the \KS\ region, because only $f_1$
contributes to the prefactor which contains no FRW term and
$\bar{b}=b$. Thus the wave
function (\ref{psijoin}) has a \textit{product structure}
\begin{equation}
  \Psi_{_{\mathrm{join}}}\ =\ \Psi_{_{\mathrm{KS}}}\,\,
  \Psi_{_{\mathrm{FRW}}}
\ ,\label{product}
\end{equation}
where the \KS\ part is an exact solution of the corresponding
\WDW\ equation, whereas the FRW wave function is a WKB solution
only. It is only the WKB structure of the wave function
(\ref{psijoin}) \textit{plus} the form of the prefactor which
facilitates the product structure.  This product structure seems
to implicate that there are no correlations between the different
regions in the quantized model. One would expect this behaviour
at least in the classical limit. This conclusion however is not
valid; the separation of the different regions in Eq.
(\ref{product}) is thus an artificial one.  First, it is
invalidated by every superposition and in particular for the
regular sine solution. Second, and more important, it is not even
true for the above wave function, simply because the variable $b$
is a variable for both the \KS\ and the FRW region. That means
that expectation values for this wave function would neither give
the FRW nor the \KS\ expectation value nor a product as one would
expect if there were no correlations between the regions. The
explanation for this behaviour seems to be an insufficient
treatment of the junction conditions. While technically nothing
seems to be wrong the very act of the junction itself seems to be
improper in a quantum field theory.  Joining the regions by
demanding $a=b$ and $a_{\mathrm{m}}=b_{\mathrm{m}}$ means that
one is simultaneously fixing a variable and its conjugate
momentum, which is obviously not in accordance with a quantum
treatment.  In other words, while quantizing the spacetime in
each region one demands the junction to be undisturbed and
smooth; there are no ``quantum fluctuations'' from one region to
the other. The same comment holds of course for every joined
model as for example the Oppenheimer--Snyder model or the bubble
spacetimes. The only way to circumvent this problem at least
formally may be to transform classically to the true degrees of
freedom. For the example of a bubble spacetime this would mean
that only the bubble radius remains which is to be quantized.

\section[ ]{The symmetric initial condition for the\\ \KS\ model}
\typeout{The SIC for the KS model}

One of the most controversial issues in quantum cosmology is the
question of how to pose a boundary condition at the wave
function. The first proposal is due to DeWitt who proposed in his
1967 paper on canonical quantum gravity \cite{dewitt_67i} a
vanishing wave function for vanishing volume. In the 80th then
the \Nob\ condition of Hartle and Hawking
\cite{hartle_hawking_83} caused some new interest in quantum
cosmology which was followed by several other suggestions of how
the boundary conditions should look like.  Here I restrict myself
mostly to the \underline symmetric \underline initial \underline
condition (SIC) \cite{conradi_zeh_91,conradi_92b} but compare the
SIC with the \Nob\ condition.

The \Nob\ condition of Hartle and Hawking is an initial condition
for paths although it may be that it makes sense for classical
trajectories only. This point of view was for example expressed
by Hawking in his comment to Zeh's contribution at the Huelva
conference on the arrow of time \cite{huelva_93}. At least for
practical purposes it is used mostly as a means for defining
initial conditions for the Hamilton--Jacobi equation or
alternatively for the classical trajectories.  In contrast to the
\Nob\ condition the SIC is an initial condition for the \WDW\
equation. The main underlying idea of the SIC is to demand the
wave function to be initially as ``simple'' as possible in
\textit{configuration space}. That this boundary condition may be
sensibly imposed is confirmed by the structure of the potential
in the \WDW\ equation which goes to zero for $\Alpha$. For the
FRW model with
arbitrary matter the potential even has the form
\begin{equation}
  V(\alpha,\{\beta_i\})\ \to\ V(\alpha)\ \to \ 0 \for \Alpha
\ .\label{pot}
\end{equation}
Here and in the following $\alpha$ is the timelike variable in
configuration space. $\alpha$ can always be chosen to be a
function of the volume $v$ of the 3--geometry and is here and in
the following defined by $\alpha=\frac 13\ln v$. The $\beta_i$
characterize all the spacelike variables like matter fields or
gravitational variables for more complicated models than the FRW
model. In view of the behavior of the potential (\ref{pot}), a
first
formulation of the SIC may therefore be
\begin{equation}
  \Psi(\alpha,\{\beta_i\})\ \to\ \Psi(\alpha)\ \to\ const
  \for\Alpha
\ .\label{sic1}
\end{equation}
But what does it mean that the wave function behaves like a
constant? The obvious choice
\begin{equation}
  \partial_{\beta_i}\Psi(\alpha,\{\beta_i\})\ \to\ 0
  \for\Alpha
\label{sic2}
\end{equation}
is ambiguous since it does not specify how (\ref{sic2}) has to be fulfilled.
For the example of oscillator eigenfunctions with
$\omega\propto e^{3\alpha}$, as in the FRW model with scalar field, all
superpositions of these eigenfunctions would satisfy this criterion.
Furthermore, it cannot be generalized for non--FRW potentials. In
order to decide whether a wave function is constant or not one needs a
basis which is here canonically given by the ``spacelike
Hamiltonian'' $H_{_{\mathrm{sp}}}$ defined by
\begin{equation}
  \partial_{\alpha}^2\Psi \ =\  H_{_{\mathrm{sp}}}\Psi
\ .\label{spham}
\end{equation}
$H_{_{\mathrm{sp}}}$ is called spacelike Hamiltonian since it
contains only spacelike momenta and is thus a Schr\"odinger like
operator. Its eigenfunctions
\begin{equation}
  H_{_{\mathrm{sp}}}h_n \ =\  E_nh_n
\label{eigen}
\end{equation}
then define canonically what a ``simple'' wave function is
supposed to mean:
\begin{equation}
  \int \Psi h_n \prod_id{\beta}_i\ \to\
  \int const\cdot h_n \prod_id{\beta}_i
  \for \alpha\to -\infty
\ .\label{sic3}
\end{equation}
This initial condition for the wave function is called \underline
symmetric \underline initial \underline condition (SIC) in order
to emphasize the character of the wave function as a particular
superposition of \textbf{all} eigenfunctions. One can simplify
this condition by using a more intuitive approach: A wave
function is constant if it is broader than
all the eigenfunctions
\begin{equation}
  \frac{b_{\Psi}}{b_{h_n}}\ \to\ +\infty \for \Alpha
\ ,\label{sic4}
\end{equation}
where $b_{\Psi}$ and $b_{h_n}$ are the widths of the wave
function and the eigenfunctions, respectively.

The definitions (\ref{sic3}) and (\ref{sic4}) of the SIC suppose
that the eigenfunctions $h_n$ are normalized in the spacelike
degrees of freedom. This seems to be a reasonable condition which
is indeed fulfilled by some important cosmological models, as
e.~g. the FRW model with arbitrary matter, Bianchi--\neun\ or the
perturbed FRW model. The \KS\ model, however, admits an unbounded
potential and does not possess normalizable eigenfunctions (see
Eq. (\ref{kspot}) below).

Before addressing the problem of the \KS\ model, I will first
review models which are better tractable by considering only
those potentials which behave as in (\ref{pot})
\cite{conradi_92b}.  One can then show that with respect to these
degrees of freedom, the condition (\ref{sic2}) is a
\emph{consequence} of the SIC as defined in (\ref{sic3}).
Consider, for example, the \WDW\ equation for the FRW model with
a scalar field
\begin{equation}
  H\Psi\ =\ \left[\partial_{\alpha}^2-3\partial_{\phi}^2 -
  9\ca{I}_1{}^2\left(
  e^{4\alpha} - \frac 13 m^2\phi^2e^{6\alpha}
  \right)\right]\Psi\ =\ 0
\ ,\label{frwphi}
\end{equation}
where $a=e^{\alpha}$ is the FRW scale factor. The uncommon
prefactors of the terms containing the scalar field result from a
reparametrization of the scalar field suited for the spherical
symmetric model but different from the usual FRW
reparametrization $\phi=\frac{1}{\sqrt{3}}\phi_{_{FRW}}$.  The
partial differential equation (\ref{frwphi}) then reduces to a
one dimensional differential equation
\begin{equation}
  H\Psi\ \approx\ \left[\partial_{\alpha}^2 - 9\ca{I}_1{}^2\left(
  e^{4\alpha} - \mbox{ $\frac 13$}m^2\phi^2e^{6\alpha}
  \right)\right]\Psi\ =\ 0
  \for\Alpha
\label{adfrw}
\end{equation}
which can be analyzed with the common methods of one--dimensional
quantum mechanics. Since the spacelike Hamiltonian
$H_{_\mathrm{{sp}}}$ has the form of a harmonic oscillator, the
spacelike eigenfunctions are known and the discussion of the
condition (\ref{sic3}) or (\ref{sic4}) can be carried through. It
turns out that it is convenient for several reasons to
introduce a repulsive potential in the Planck era, as e.~g.
\begin{equation}
  V_{_{\mathrm{P}}}\ =\ -C^2e^{-2\alpha}
\label{planckpot}
\end{equation}
with the arbitrary constant $C^2$. The WKB approximation can only
then be used for $\Alpha$ and the solutions behave exponentially.
This in turn facilitates the possibility of selecting one
solution by a normalization condition. Furthermore the SIC in its
strong formulation (\ref{sic3}) can be fulfilled.  The Planck
potential seems to be artificial, however, since the physics at
the Planck scale is entirely unknown and quantum gravity is often
regarded as the low energy limit of some GUT, a Planck potential
may arise as an effective potential of such a theory (the
effective axion potential, e.~g., almost leads to a Planck
potential). Furthermore, one can get completely rid of the Planck
potential by regarding it solely as a regulator which gives one a
unique wave function.

With the Planck potential the FRW model with massive scalar field
can easily be solved (in the considered regions)
\cite{conradi_zeh_91,conradi_92b}. For this model and for the
perturbed FRW model, the SIC approximately leads to the same wave
functions as the \Nob\ condition
\cite{hartle_hawking_83,hawking_84}. This is to be expected since
in both examples the condition (\ref{sic2}) for matter fields has
been used by either boundary condition. However, there is an
important difference: While the SIC leads to a solution of the
\WDW\ equation, this is not so for the \Nob\ condition: It has
often been emphasized after the 1985 debate between Page and
Hawking \cite{page_85,hawking_85} that the condition $\phi=const$
is valid only for the beginning of the classical trajectories but
not for the end and that therefore the wave function possesses an
additional oscillating term which reflects this behaviour. This
means that the condition $\Psi\to1$ is regarded not as an initial
condition for the \WDW\ equation but as an initial condition for
the Hamilton--Jacobi equation.  As more explicitly argued in the
conclusion it is exactly this difference which leads to the
disagreement for the arrow of time derived with both boundary
conditions, for a more detailed discussion see e.~g.
\cite{conradi_zeh_91,hawking_85,zeh_book,zeh_94a,hawking_etal_93}
and the literature cited therein. For a critique of using
semiclassical methods only, see in particular
\cite{kiefer_zeh_94}.

In the context of this paper I am only
interested in the dust case where the \WDW\ equation is given by
\begin{equation}
  H\Psi\ =\ \left[\partial_{\alpha}^2 - C^2e^{-2\alpha} -
  9\ca{I}_1{}^2\left(
  e^{4\alpha} - a_{\mathrm{m}}e^{3\alpha}\right)\right]\Psi\ =\ 0
\ . \label{dustfrw}
\end{equation}
By generalizing the SIC wave function from the scalar field model
one finds
\begin{equation}
\hspace*{-1mm}
  \Psi\ =\ \exp \left\{ 3i\ca{I}_1\left(
  \frac{a^2}{4}\left(2-\frac{a_{\mathrm{m}}}{a}\right)
  \sqrt{\frac{a_{\mathrm{m}}}{a}-1} -
  \frac{a_{\mathrm{m}}^2}{8}\arcsin\left(\frac{a_{\mathrm{m}}-2a}
  {a_{\mathrm{m}}}\right)
  \right)\right\}
\label{frwdust}
\end{equation}
($a=e^{\alpha}$) in the region, where the Planck potential can be
neglected.  Strictly speaking, however, it is not possible to
determine a SIC wave function for the FRW--dust model since the
spacelike Hamiltonian vanishes.

As mentioned above, the SIC (\ref{sic3}) is not directly
applicable to the
\KS\ model, whose \WDW\ equation reads
\begin{eqnarray}
  \Bigg[ z^2\partial^2_z + \mbox{$\frac 12$}z\partial_z -
  2zb\partial_b\partial_z + \partial^2_{\phi} +
  \ca{I}^2zb^2 \left(
  z - \Lambda zb^2 - zb^2m^2\phi^2 \right)\Bigg] \Psi \ =\ 0
\\
  \Longleftrightarrow \qquad \left[
  \partial_{\alpha}^2 - \partial_{\sigma}^2 - 3\partial_{\phi}^2
  + 3\ca{I}^2 \left(
  - e^{4\alpha + 2\sigma} + e^{6\alpha}(\Lambda + m^2\phi^2) \right)
  \right]\Psi\ =\ 0
\ .\label{kspot}
\end{eqnarray}
The \WDW\ equation is here displayed with a scalar field, and the
Laplace--Beltrami factor ordering is chosen (for the factor
ordering the scalar field is neglected).  Again $\alpha:=\frac
13\ln zb^2$ is the timelike variable and $\sigma := \frac 13\ln(z/b)$
and $\phi$
are spacelike. While the scalar field poses no problem (its
potential is bounded; and since (\ref{sic2}) can be used for
$\Alpha$, it is equivalent to a cosmological constant and
therefore not considered in the following), the potential for
$\sigma$ is unbounded for $\sigma\to\infty$. It is therefore
impossible to find normalizable eigenfunctions for the spacelike
Hamiltonian.  Condition (\ref{sic2}), too, cannot be used because
the $\sigma$--dependence of the potential does not vanish for
$\Alpha$.

As a first attempt to apply the SIC in spite of this difficulty
one can investigate the effects of a ``Planck potential'' for the
\KS\ model.  The Planck potential is given by exploiting the fact
that the dynamics for $b$ in the \KS\ model should be identical
to the dynamics of $a$ in the FRW model, regardless of a Planck
potential. Using the ansatz (\ref{planckpot}) for the FRW Planck
potential one gets
\begin{equation}
  V_{_{\mathrm{P}}} = - 10C^2\frac{z}{b^6} + \tilde{V}(b)
\ ,\label{ksplanck}
\end{equation}
as a Planck potential for the \KS\ model. Here $\tilde{V}(b)$ is
an arbitrary function which for simplicity is set to
$\tilde{V}=D^2/b^6$ in the following. Another ansatz for the FRW
Planck potential or a general $\tilde{V}$ does not alter the
conclusions. The additional Planck potential does not render the
potential bounded in $\sigma$, since in the \{$\alpha$,
$\sigma$\} representation it has the form
\begin{equation}
  \frac{1}{3\ca{I}^2} V_{_{\mathrm{P}}}=10C^2e^{-2\alpha+8\sigma} -
  D^2e^{-3\alpha+6\sigma}
\ .
\end{equation}
Obviously, the potential is still unbounded.

However, one can use the FRW model with its SIC wave function in
another way by considering the joined model. The FRW part of the
wave function fixes the \KS\ part of the wave function by the
product structure of the solution. One therefore has to analyze
the joined model with a Planck potential included for each
region. It turns out that again this model has a potential of the
form
\begin{equation}
  V = \ca{I}_2{}^2zb^2 \left( zf_1(b) + f_2(b)  \right)
\end{equation}
with
\begin{eqnarray}
  f_1(b) &=& 1 - \Lambda b^2 - 10C^2b^{-6}
\nonumber \\
  f_2(b) &=& -\left(z_{\mathrm{m}} + 3\mbox{$\frac{\ca{I}_1}{\ca{I}_2}$}
  b_{\mathrm{m}}\right) +
  2\mbox{$\frac{\ca{I}_1}{\ca{I}_2}$} b + D^2b^{-6}
  -13\mbox{$\frac{\ca{I}_1}{\ca{I}_2}$}C^2b^{-5}
\ .
\end{eqnarray}
The joined model with a Planck potential is thus exactly solvable
and its solution is given by (\ref{psijoin}). The SIC wave
function for the FRW model in the region where the Planck
potential can be neglected is given by (\ref{frwdust}). This is
identical with the FRW part of the joined model as given by
(\ref{hjjoin}). The \KS\ part of the wave function can thus be
interpreted as fulfilling the SIC. For the \textit{vacuum} FRW
and \KS\ model and in the region where the Planck potential
can be neglected, the wave function simplifies to
\begin{equation}
  \Psi\ =\ \frac{1}{\sqrt{\frac 13\Lambda b^3 - b}}
  \exp\left\{ \pm i\ca{I}_2\left( zb\sqrt{\frac 13\Lambda b^2 - 1}
  \right) \right\}
\ .\label{kssic}
\end{equation}
Except for the prefactor this is the wave function derived by
Laflamme and Shellard in order to fulfill the \Nob\ condition
\cite{laflamme_shellard_87}. As emphasized in particular by
Halliwell and Louko \cite{halliwell_louko_ii} a more
sophisticated treatment of the \Nob\ condition shows that due to
a freedom in choosing a ``contour of integration'' one could get
either one of the signs in the exponent or the sine or the cosine
as superpositions of these solutions. Only an additional choice
restricts one to the exponential solution again. But while
Halliwell and Louko discuss the vacuum model only, Laflamme and
Shellard are primarily interested in the model with scalar field.
In this case the solution is derived by assuming $\phi=const$
initially; that is the scalar field behaves as a cosmological
constant. It is important in their approach that a
\textit{classical} scalar field behaves like a constant only for
one end of the evolution. Since the \Nob\ wave function is given
by the exponent of the action evaluated at the classical
trajectory the above solution is argued to be valid ``initially''
only.  It is thus interpreted as a solution of the
Hamilton--Jacobi equation while the corresponding solution of the
\WDW\ equation would admit an additional oscillating term.

\section{Conclusion}
\typeout{Conclusion}

By reviewing the classical analysis of the \KS\ model, I have
emphasized the pathological character of this model: The
incompleteness of the model, the possibility of an inhomogeneous
$z$ and the unboundedness of its spacelike potential. It is,
nevertheless, useful as a toy model since it is the most simple
anisotropic model and the joined model is even inhomogeneous. The
more realistic Bianchi--\neun\ model is far more complicated and
an analytical treatment has so far not been given.

In Sec.~3 the \WDW\ equation was solved exactly for a whole class
of \KS\ like models and a complete set of wave functions was
given.  The treatment of the general spherically symmetric model
served here as a guide. The class of solved models includes the
\KS\ model with a cosmological constant and dust as well as the
joined model. The derived exact solutions
are of WKB form
\begin{equation}
  \Psi_{_M}\ =\ De^{\pm iS_M}
\ ,\label{exact}
\end{equation}
where $S_M$ are solutions of the Hamilton--Jacobi solution, and
$D$ is the WKB prefactor.  This is somewhat surprising and one
therefore has to be careful when basing conclusions on the
breakdown of the WKB approximation as in \cite{nakamura_etal_93}.
Compared to the other sets of solutions given below, the
$\Psi_{_M}$ have the advantage of great applicability. A
cosmological constant and dust and even the joined model are
contained in the class of solvable models.

For the vacuum model an alternative set of exact WKB solutions
$\Psi_K= e^{iS_K}=e^{iS_K}$ (with constant prefactor) is known
which is given for the vacuum model by
\begin{equation}
  \Psi_{_K}\ =\ \exp\left\{ \pm \frac{i\ca{I}}{2} \left(
  -Kz^2b -  \frac bK \right) \right\}\ =\ e^{\pm iS_K}
\ ,\label{exactK}
\end{equation}
where $K$ is an arbitrary (complex) constant
\cite{louko_vachaspati_89,uglum_92}.  This solution is found by
transforming the original \WDW\ equation (\ref{wdw}) into a
Klein--Gordon equation by $u = (\ca{I}/4)z^2b$ and $v = \ca{I}b$.
The transformation can be generalized to an arbitrary $f_1(b)$,
but a generalization to a nonvanishing $f_2(b)$ has not yet been
found.  This set of solutions is complementary to the one
discussed in this paper in the following sense: $S_M$ and $S_K$
both solve the Hamilton--Jacobi equation but while $S_M$
represents the classical solutions with a fixed constant of
integration $2M = b_{\mathrm{m}}$ for $b(t)$, $S_K$ represents
the classical solutions with a fixed constant of integration $K$
for $z(t)$, c.~f. Eq.~(\ref{classical}). It is worth noting,
however, that all solutions corresponding to $S_K$ are identical
in the vacuum model.

Another set of solutions for the vacuum model was first given by
Ryan \cite{ryan_72} by separating the \WDW\ equation by the
transformation $x=z$ and $y=zb$. For a certain factor ordering
one then gets the solutions
\begin{equation}
  \Psi_{_k}\ =\
  z^{\pm ik}K_{ik}\left(zb\right)
\ ,\label{exactk}
\end{equation}
where $k$ is an arbitrary constant of separation and
$K_{ik}\left(zb\right)$ is the MacDonald function (the
MacDonald function as solution of a modified Bessel equation has
been chosen since it vanishes exponentially for great values of
$y=zb$).  Unlike the other sets of solutions these functions do
not have a WKB structure.  These solutions can be generalized to
a rather arbitrary factor ordering and even for some arbitrary
functions in the potentials. Unfortunately however, neither a
cosmological constant nor the dust case fits into these class of
models, so that is usage seems to be restricted to the vacuum
model and the coupling of a massless scalar field.

The detailed analysis of how the different set of solutions are
related requires a well--defined scalar product for this model.
This is an open problem at the moment.

The joined model which consist of a \KS\ cylinder inserted
between two FRW half--spheres was investigated in detail in
Sec.~4.  It is mainly the WKB structure of the wave function
which allows the unique solution of this model to be written as a
product of solutions of the respective spacetime regions. In
spite of the product structure the solution can not be
interpreted as admitting no correlations between the different
spacetime regions since the variable $b$ is defined for the whole
spacetime. This reflects the fact that although one has quantized
the spacetime in the different regions one does not allow
fluctuations between them --- the junction is treated
classically.

One can nevertheless use the wave function for the joined model with
its product structure in order to determine a wave function for the
pure \KS\ model which satisfies the \underline symmetric \underline
initial \underline condition (SIC).  This initial condition cannot be
directly applied because of the structure of the potential in the
\WDW\ equation. Neither does it simplify to a pure function of the
volume for small volumes nor does it admit normalizable spacelike
eigenfunction due to the unboundedness of the spacelike potential like
other reasonable models as the Bianchi--\neun\ model. The SIC wave
function for the \KS\ model is then determined by the \KS\ part of the
product wave function since its FRW part is the respective SIC wave
function.  It turns out that the wave function which satisfies the SIC
is approximately the same as the one which was given by Laflamme and
Shellard for the \Nob\ condition \cite{laflamme_shellard_87}.

The interpretation of the solution is, however, different for the
two boundary conditions. In case of the \Nob\ condition the wave
function is regarded as a solution of the Hamilton--Jacobi
equation. This is due to the fact that only classical
trajectories were considered and that the condition of
$\phi=const$ can only be satisfied at one end of the classical
solution while at the other one $\phi$ behaves in a more
complicated way. (This is a special case of the general
observation that a spacetime which ``starts'' regular will
develop inhomogeneities and anisotropies and thus end up
irregular. The beginning respective the end is thereby
\textit{defined} by the homogeneity of the state.) Since the
\Nob\ wave function is given by the exponent of the classical
action Laflamme and Shellard concluded that the approximation and
thus the wave function is good only ``initially''.  This means
that their wave function can only be regarded as a solution of
the Hamilton--Jacobi equation while in this interpretation a
solution of the \WDW\ equation would admit an additional
(oscillating) factor which corresponds to the complicated ``end''
of the classical trajectory.  Since they argue that their wave
function admits no initial anisotropies its interpretation as a
solution of the Hamilton--Jacobi equation means that the wave
function predicts a universe which ends up anisotropic.

The SIC on the other hand is interpreted as an initial condition
for the solutions of the \WDW\ equation.  This marks a decisive
difference because of the dynamical structure of this equation.
The \WDW\ equation is a wave equation (the signature of its
kinetic term admits one minus sign, the rest are plus signs)
which thus admits one internal timelike variable. This timelike
variable, the so called intrinsic time, has nothing to do a
priori with the physical time but is that variable with respect
to which the dynamics of the wave is defined.  The timelike
variable can always be chosen to be the volume of the
three--geometry. Since the dynamics is a dynamics with respect to
the volume it follows directly that a prediction of a homogeneous
and isotropic state for small volumes is irrespective of an
expanding or a contracting universe. Moreover, even the very
notion of expanding and contracting does not make sense in
quantum cosmology due to the absence of an external time
parameter; one can only conclude how a variable changes with
varying volume.  Since the SIC arrives at approximately the ``no
boundary'' wave function which is thought to predict an isotropic
initial state (or more generally a homogeneous and isotropic
initial state) it can be concluded that the wave function
predicts an isotropic state for small volumes. The universe in
this interpretation is thus predicted to be isotropic at its
``beginning'' and at its ``end''.

The different interpretations of the wave function lie at the heart of
the ongoing controversy on the symmetry or asymmetry of the arrow of
time. Since the ``no boundary'' wave function is interpreted as
Hamilton--Jacobi function it is thus consequent that with this
assumptions an asymmetric arrow of time is derived (for a critique of
the consistency of using semiclassical methods only in quantum
cosmology see \cite{kiefer_zeh_94}) while the SIC as an initial
condition for the \WDW\ equation arrives at an symmetric arrow of
time. A symmetric arrow of time does not lead to paradoxes in the
recollapsing part of the universe since it is not possible to build a
wave function which represents a recollapsing universe: Quantum
interference effects are dominant at the turning point, see
\cite{zeh_94a} for a more detailed argument. This is a direct
consequence of the above considerations but can be explicitly verified
for some models too.

The exact solutions (\ref{exact}) were in this paper mainly used
in order to find a wave function which satisfies the SIC and then
to confirm that the SIC leads to an isotropic beginning of the
classical universe.  There are, of course, other open questions
in this class of models for which the exact solutions may be
helpful. I will mention but two of them. First, what is the right
scalar product? The given wave function for example have either a
apparent singularity at the classical borderline or the wave
function diverges exponentially for great $b$. In neither case it
seems to be appropriate simply to interprete the wave function as
the probability for a some configuration. Second, one should
include a more realistic matter source like a massive scalar
field. This is in particular important in order to analyze the
behavior of the wave function at the classical turning point:
Does there exist a wave function which resembles a recollapsing
universe? For the FRW model Kiefer was able to show that this is
impossible \cite{kiefer_88b}. Does the additional degree of
freedom change this behaviour by avoiding the interference of the
``expanding'' and ``recollapsing'' part of the wave packet due to
the extra dimension of the configuration space? In order to
answer this question, one already needs a well defined scalar
product.

\enlargethispage{\baselineskip}

\vspace{15mm}
\noindent
{\Large\bfseries Acknowledgment}

\

I wish to thank H.--Dieter Zeh for many invaluable discussions
and for reading and commenting on a preliminary version of this
paper.  Furthermore, I thank Stefan Kehrein, Karel Kucha\v{r} and
in particular Claus Kiefer whose help greatly improved the
manuscript.

\end{document}